\documentclass{ws-p8-50x6-00}

\begin{document}

\title{Vector meson photoproduction in the quark model}

\author{Q. Zhao}

\address{Department of Physics, University of Surrey, Guildford, GU2 7XH, UK
\\E-mail: qiang.zhao@surrey.ac.uk}


\maketitle

\abstracts{
We present a quark model study of 
the $\omega$ meson photoproduction near threshold. 
With a limited number of parameters, all the data 
in history are reproduced. The roles played by the {\it s}-
and {\it u}-channel processes (resonance excitations and nucleon
pole terms), 
as well as the {\it t}-channel
{\it natural} (Pomeron) and {\it unnatural} parity (pion) exchanges 
are clarified.
This approach provides a framework 
for systematic study of vector meson photoproduction
near threshold. 
}

\section{Introduction}

For a long time, the experimental study of the neutral vector 
meson ($\omega$, $\rho^0$ and $\phi$) photoproduction 
was concentrated on 
high energy regions, where the diffractive 
process played a dominant role and 
could be accounted for by 
a soft Pomeron exchange model.  
Recently, the availabilities of high intensitve 
electron and photon beams at JLAB, ELSA, ESRF, and 
SPring-8 give accesses to excite nucleons
with the clean electromagnetic probes.
Thus, vector meson production via resonance 
excitations provides an ideal tool to study 
the non-diffractive mechanisms in vector meson 
photoproduction near threshold. 
Concerning the resonance excitations in this 
reaction, the other essential motivation
is to search for ``missing resonances", 
which were predicted by the nonrelativistic constituent
quark model (NRCQM)~\cite{NRCQM-1}, 
but not found in $\pi N$ scatterings.
Vector meson photoproduction near threshold
might provide supplementary knowledge of those missing
resonances and their couplings to vector mesons.

In this proceeding, a quark model
approach to vector meson photoproduction
near threshold is applied to the $\omega$ meson
photoproduction. 
Our purpose is to provide 
a framework on which a systematic study of 
resonance excitations becomes possible.
Our model
consists of three processes: (i) {\it s}- and {\it u}-channel
vector meson production with an effective Lagrangian (S+U);
(ii) {\it t}-channel Pomeron exchange ($\cal{P}$) for $\omega$,
$\rho^0$ and $\phi$ production~\cite{Donnachie}; 
(iii) {\it t}-channel
light meson exchange. 
Namely, in the $\omega$ meson photoproduction, the $\pi^0$ exchange
is taken into account. 

In the $SU(6)\otimes O(3)$ symmetry limit, 
the constituent quark $\psi$ couples to a vector meson $\phi^\mu_m$ 
via an effective Lagrangian~\cite{zhao-98,zhao-omega-2001}: 
\begin{equation} \label{Lagrangian}  
L_{eff}=\overline{\psi}(a\gamma_\mu +  
\frac{ib\sigma_{\mu\nu}q^\nu}{2m_q}) \phi^\mu_m \psi,
\end{equation}  
where $a$ and $b$ are the overall parameters introduced
at quark level for baryon states. 
In this way, all the {\it s}- and {\it u}-channel 
resonances and nucleon pole terms can be included. 
The {\it t}-channel vector meson exchange and 
contact term from the Lagrangian 
will only contribute in the charged 
vector meson photoproduction. 
In $\gamma p \to \omega p$,  
eight low-lying resonances: 
$P_{11}(1440)$, $S_{11}(1535)$, $D_{13}(1520)$, 
$P_{13}(1720)$, $F_{15}(1680)$, $P_{11}(1710)$, $P_{13}(1900)$, and
$F_{15}(2000)$,
with quark harmonic oscillator shell $n\le 2$
are explicitly included~\cite{PDG2000}, 
while higher mass states are treated degenerate with $n$. 
We refer the readers to Ref.~\cite{zhao-98,zhao-omega-2001}
for details of this approach.

%%%%%%%%%%%%%%%%%%%%%%%% FIG 1 and FIG 2
%\vspace*{-1.5cm}
\begin{figure}[htb]
%\begin{center}
\begin{minipage}[t]{50mm}
\epsfxsize=14pc 
\epsfbox{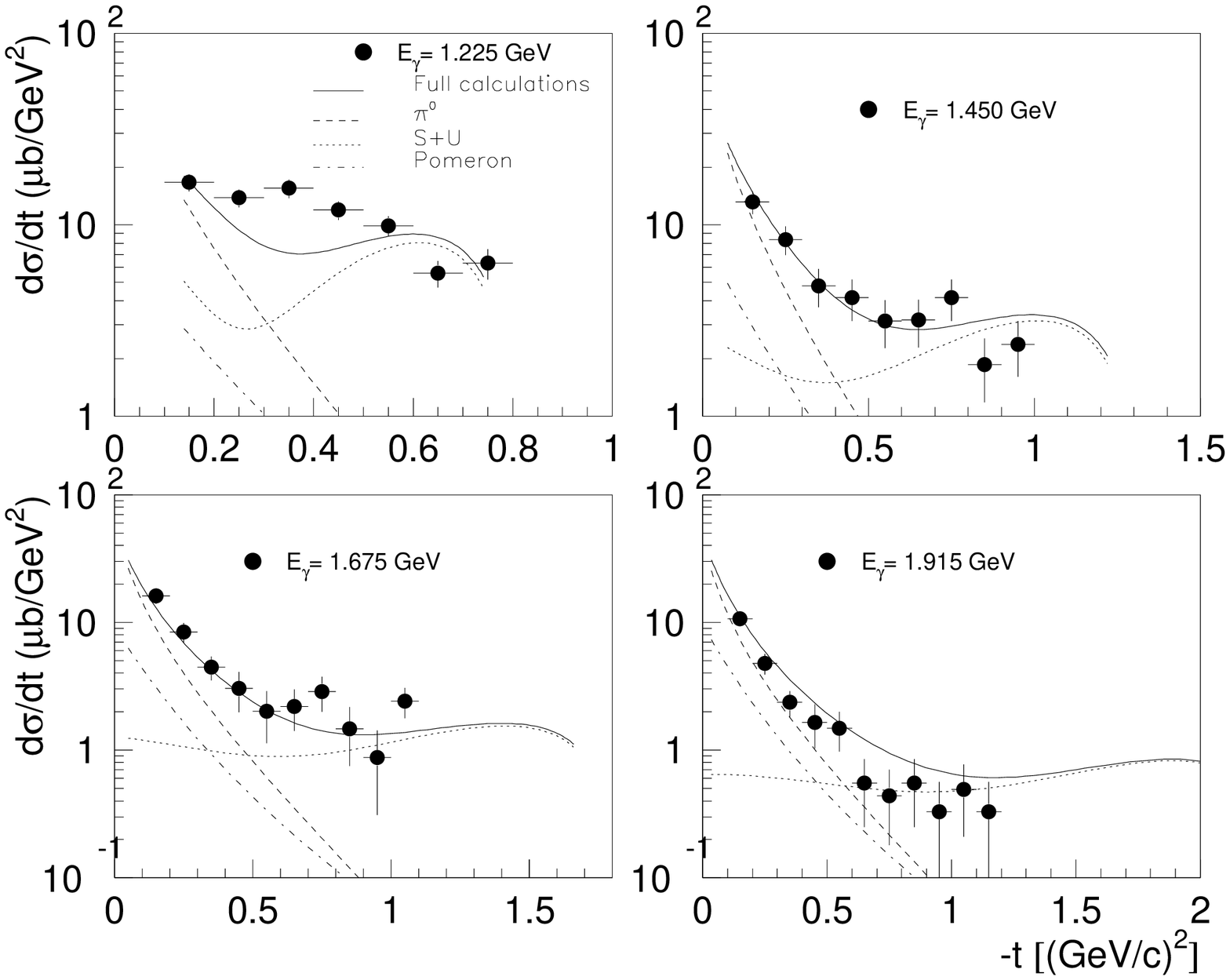} 
%\includegraphics[width=14pc]{nstar_fig1.eps}
%\framebox[79mm]{\rule[-26mm]{0mm}{52mm}}
\caption{Differential cross section for $\gamma p \to \omega p$.}
\protect\label{fig:(1)}
\end{minipage}
\hspace{\fill}
\begin{minipage}[t]{60mm}
\epsfxsize=14pc 
\epsfbox{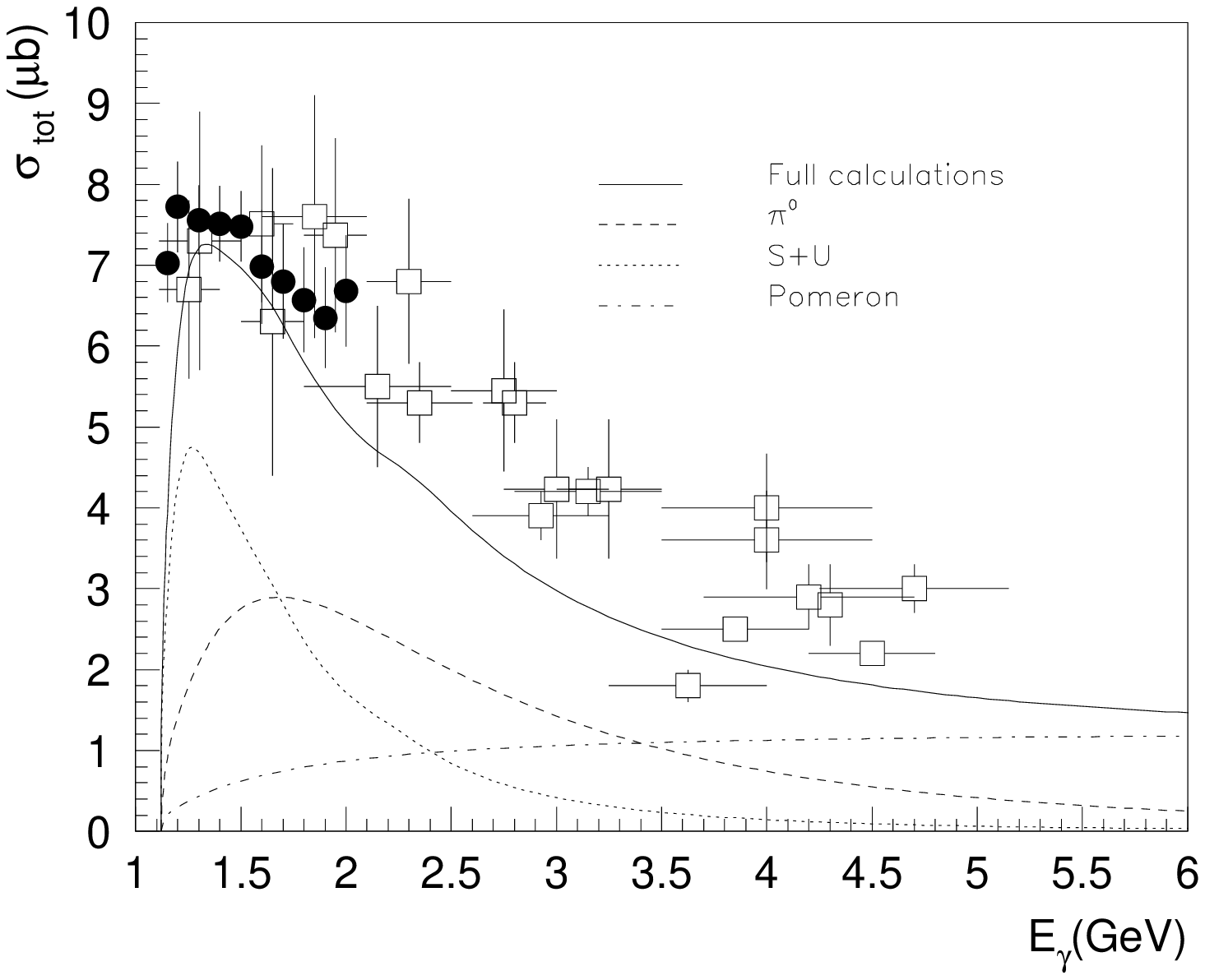} 
%\includegraphics[width=14pc]{nstar_fig2.eps}
%\framebox[74mm]{\rule[-26mm]{0mm}{52mm}}
\caption{Total cross section for $\gamma p\to \omega p$.}
\protect\label{fig:(2)}
\end{minipage}
%\end{center}
\end{figure}
%%%%%%%%%%%%%%%%%%%%%%%%%

\vspace*{-0.5cm}
\section{Analysis}

In Fig.~\ref{fig:(1)}, differential 
cross sections for four energy bins, $E_\gamma=$1.225,  
1.450, 1.675 and 1.915 GeV, are presented and compared 
with the SAPHIR data~\cite{klein}. 
Results for exclusive processes are also presented. 
It shows that near threshold the $\pi^0$ exchange
plays a dominant role over the other two processes,
in particular at small angles. With the increasing energy
the {\it natural} parity Pomeron exchange 
becomes more and more important which will produce 
interesting interfering effects in polarization observables.
In Ref.~\cite{zhao-omega-2001} we show that the 
{\it natural} and {\it unnatural} parity exchanges
can be constrained well by the measurement
of forward angle parity asymmetries~\cite{ballam73}.
The dotted curves in Fig.~\ref{fig:(1)} represent 
the exclusive calculations of the {\it s}- and {\it u}-channel
processes. They account for the large angle behavior 
in the differential cross sections, but 
have only a small impact on the small-angle cross sections.
This feature justifies the method we constrain
the {\it t}-channel processes.

%%%%%%%%%%%%%%%%%%%%%%%% FIG 3 
%\vspace*{-1.5cm}

\begin{figure}[t]
%\figurebox{12pc}{6pc}{nstar_fig3.eps} 
\epsfxsize=15pc 
\epsfbox{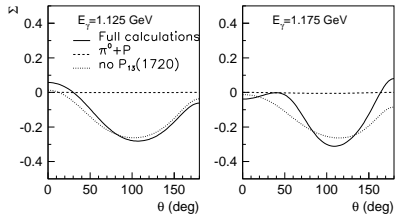} 
\caption{Polarized beam asymmetry.  See the text for notations. 
\protect\label{fig:(3)}}
\end{figure}
%%%%%%%%

In Fig.~\ref{fig:(2)}, the total cross section 
for $\gamma p \to \omega p$ is 
given. 
Interestingly, it shows that the dominant process
near threshold is the S+U, 
although it falls down rapidly with the increasing
energy. The Pomeron exchange becomes dominant over 
the pion exchange above $E_\gamma\sim 3.5$ GeV, which 
is consistent with the experimental results~\cite{ballam73}.
Note that in Ref.~\cite{oh}, a phenomenological model
with parameters predicted by the $^3P_0$ quark-pair-creation
model~\cite{capstick} 
obtains quite different results for exclusive processes.

For the purpose of investigating 
individual resonance excitations, we have to turn to polarization
observables in which small effects from individual resonances 
might be picked up.
Following the convention of Ref.~\cite{tabakin}, 
in Fig.~\ref{fig:(3)} we present predictions for 
$\Sigma\equiv (2\rho^1_{11}+\rho^1_{00})/(2\rho^0_{11}+\rho^0_{00})$,
where $\rho^1$ and $\rho^0$ are density matrix elements in the helicity 
space~\cite{schilling}.
One of the most important features of $\Sigma$ is that 
large asymmetries cannot be produced by the $\cal{P}$ or $\cal{P}$+$\pi^0$.
As shown by the solid curves, large asymmetries are produced by
the interferences between the $\cal{P}$+$\pi^0$ and 
S+U processes.

In this study, we find that the $P_{13}(1720)$ and $F_{15}(1680)$, 
which are classified to $[56, ^2 8, 2,2, J]$ 
quark model representation, play a strong role in this reaction. 
The $S_{11}(1535)$ and $D_{13}(1520)$ of $[70, ^2 8, 1,1, J]$ 
have also relatively large effects. In Fig.~\ref{fig:(3)},
we show that without the $P_{13}(1720)$, the asymmetry 
will be significantly changed (see the dotted curves).
The predictions can be compared to the preliminary 
data from GRAAL collaboration~\cite{graal}.

Another interesting observable with polarized photon beam is,
$\Sigma_A\equiv (\rho^1_{11}+\rho^1_{1-1})/(\rho^0_{11}+\rho^0_{1-1})
=(\sigma_\parallel-\sigma_\perp)/(\sigma_\parallel+\sigma_\perp)$,
where $\sigma_\parallel$ and $\sigma_\perp$ represent the pion 
decay cross sections of the $\omega$ meson with the pions 
submerged in or perpendicular to the photon polarization plane.
As found in Ref.~\cite{zhao-omega-2001}, this observable 
is more sensitive to small contributions from individual resonances.

Summarily,
we present a quark model study of vector meson photoproduction
near threshold by applying it to $\gamma p \to \omega p$. 
It provides a framework on which a systematic 
investigation of resonance excitations in vector meson production
can be done. 
Our predictions should be confronted 
with the forthcoming data from GRAAL and JLAB 
in the near future.

\section*{Acknowledgments}
Fruitful discussions with E. Hourany concerning the GRAAL 
experiment are acknowledged.

\end{document}